\documentclass[aps,prb,twocolumn,superscriptaddress,notitlepage]{revtex4-1}
\usepackage{graphicx}
\usepackage[caption=false]{subfig}
\usepackage{float}
\usepackage{natbib}
\usepackage{xcolor}
\usepackage{xr}
\usepackage{amsmath}
\usepackage{amssymb}
\usepackage{array}
\usepackage{wasysym}
\usepackage{color,soul}
\usepackage{braket}
\usepackage{verbatim}
\usepackage{hyperref}
\usepackage{gensymb}
\usepackage{url}
\usepackage{dirtytalk}

\usepackage{filecontents}
\begin{document}

\title{Dome-Shaped Superconducting Phase Diagram Linked to Charge Order in LaRu$_{3}$Si$_{2}$}

\date{\today}
\author{KeYuan Ma}
\thanks{These authors contributed equally to experiments.}
\affiliation{Max Planck Institute for Chemical Physics of Solids, 01187 Dresden, Germany}

\author{I.~Plokhikh}
\thanks{These authors contributed equally to experiments.}
\affiliation{PSI Center for Neutron and Muon Sciences CNM, 5232 Villigen PSI, Switzerland}

\author{J.N. Graham}
\thanks{These authors contributed equally to experiments.}
\affiliation{PSI Center for Neutron and Muon Sciences CNM, 5232 Villigen PSI, Switzerland}

\author{C. Mielke III}
\thanks{These authors contributed equally to experiments.}
\affiliation{PSI Center for Neutron and Muon Sciences CNM, 5232 Villigen PSI, Switzerland}
\affiliation{Microstructured Quantum Matter Department, Max Planck Institute for the Structure and Dynamics of Materials, Luruper Chaussee 149, 22761 Hamburg, Germany}

\author{V. Sazgari}
\thanks{These authors contributed equally to experiments.}
\affiliation{PSI Center for Neutron and Muon Sciences CNM, 5232 Villigen PSI, Switzerland}

\author{H.~Nakamura}
\affiliation{Institute for Solid State Physics (ISSP), University of Tokyo, Kashiwa, Chiba 277-8581, Japan}

\author{S.S. Islam}
\affiliation{PSI Center for Neutron and Muon Sciences CNM, 5232 Villigen PSI, Switzerland}

\author{S.~Shin}
\affiliation{PSI Center for Neutron and Muon Sciences CNM, 5232 Villigen PSI, Switzerland}

\author{P. Kr\'{a}l}
\affiliation{PSI Center for Neutron and Muon Sciences CNM, 5232 Villigen PSI, Switzerland}

\author{O. Gerguri}
\affiliation{PSI Center for Neutron and Muon Sciences CNM, 5232 Villigen PSI, Switzerland}

\author{H.~Luetkens}
\affiliation{PSI Center for Neutron and Muon Sciences CNM, 5232 Villigen PSI, Switzerland}

\author{F.O. von Rohr}
\affiliation{Department of Quantum Matter Physics, University of Geneva, CH-1211 Geneva, Switzerland}

\author{J.-X.~Yin}
\affiliation{Department of Physics, Southern University of Science and Technology, Shenzhen, Guangdong, 518055, China}

\author{E.~Pomjakushina}
\affiliation{PSI Center for Neutron and Muon Sciences CNM, 5232 Villigen PSI, Switzerland}

\author{C. Felser}
\affiliation{Max Planck Institute for Chemical Physics of Solids, 01187 Dresden, Germany}

\author{S.~Nakatsuji}
\affiliation{Institute for Solid State Physics, University of Tokyo, Kashiwa, 277-8581, Japan}

\author{B.~Wehinger}
\affiliation{European Synchrotron Radiation Facility, 71 Avenue des Martyrs, 38000 Grenoble, France}

\author{D.J.~Gawryluk}
\affiliation{PSI Center for Neutron and Muon Sciences CNM, 5232 Villigen PSI, Switzerland}

\author{S. Medvedev}
\email{sergiy.medvediev@cpfs.mpg.de} 
\affiliation{Max Planck Institute for Chemical Physics of Solids, 01187 Dresden, Germany}

\author{Z. Guguchia}
\email{zurab.guguchia@psi.ch} 
\affiliation{PSI Center for Neutron and Muon Sciences CNM, 5232 Villigen PSI, Switzerland}


\begin{abstract}
\textbf{The interplay between superconductivity and charge order is a central focus in condensed matter research, with kagome lattice systems offering unique insights. The kagome superconductor LaRu$_{3}$Si$_{2}$ ($T_{\rm c}$ ${\simeq}$ 6.5 K) exhibits a hierarchy of charge order transitions: primary ($T_{\rm co,I}$ ${\simeq}$ 400 K), secondary ($T_{\rm co,II}$ ${\simeq}$ 80 K), and an additional transition at ($T^{*}$ $\simeq$ 35 K). The transitions at  $T_{\rm co,II}$ and $T^{*}$ are linked to electronic and magnetic responses as revealed by muon-spin rotation and magnetotransport experiments. However, the connection between superconductivity, charge order, and electronic responses remains unclear. By employing magnetotransport and X-ray diffraction techniques under pressures of up to 40 GPa, we observe that $T_{\rm c}$ rises to 9 K at 2 GPa, remains nearly constant up to 12 GPa, and then decreases to 2 K at 40 GPa, resulting in a dome-shaped phase diagram. The resistivity anomaly at $T^{*}$ and magnetoresistance also exhibit a similar dome-shaped pressure dependence. Furthermore, we find that charge order transitions from long-range to short-range above 12 GPa, correlating with the suppression of $T_{\rm c}$, suggesting superconductivity is closely tied to the charge-ordered state. Specifically, $T_{\rm c}$ peaks when charge order and the normal-state electronic responses are optimized. In contrast to systems like the cuprates, transition metal dichalcogenides, and other kagome materials, where superconductivity typically competes with charge order, LaRu$_{3}$Si$_{2}$ displays a pronounced interdependence between these two phenomena. This distinctive behavior sheds new light on the connection between superconductivity and charge order, offering avenues for theoretical advancements in understanding superconductivity.}
\end{abstract}

\maketitle

\section{Introduction}

Kagome superconductors \cite{Syozi,GuguchiaNPJ,JiaxinNature,GuguchiaPRM,GuguchiaCSS,Kiesel,
GuguchiaPlokhikh,Barz,Vandenberg,BOrtiz2,BOrtiz3,QYin} have captured significant interest in condensed matter physics, offering a fertile playground for exploring various intertwined quantum phases. Notably, the $A$V$_{3}$Sb$_{5}$ family \cite{BOrtiz2,BOrtiz3,QYin} (where $A$ is an alkali metal) has demonstrated a rich array of electronic phenomena \cite{GuguchiaNPJ,JiaxinNature,YJiang,GuguchiaMielke,TNeupert,MHChristensen2022,GWagner,Grandi,GuguchiaRVS,Wang2021,KhasanovCVS,YXu,GuoMoll,SYang,MDenner,MHChristensen,Tazai2022,Balents,Nandkishore,Qimiao}, including superconductivity, charge order, time-reversal symmetry breaking, orbital order, and nematic transitions. These systems highlight the kagome lattices potential to host complex interactions, driven by its unique geometry and frustrated magnetic structure, which may stabilize exotic states of matter and encourage the interplay between multiple order parameters.

\begin{figure*}
    \centering
    \includegraphics[width=1.2\textwidth]{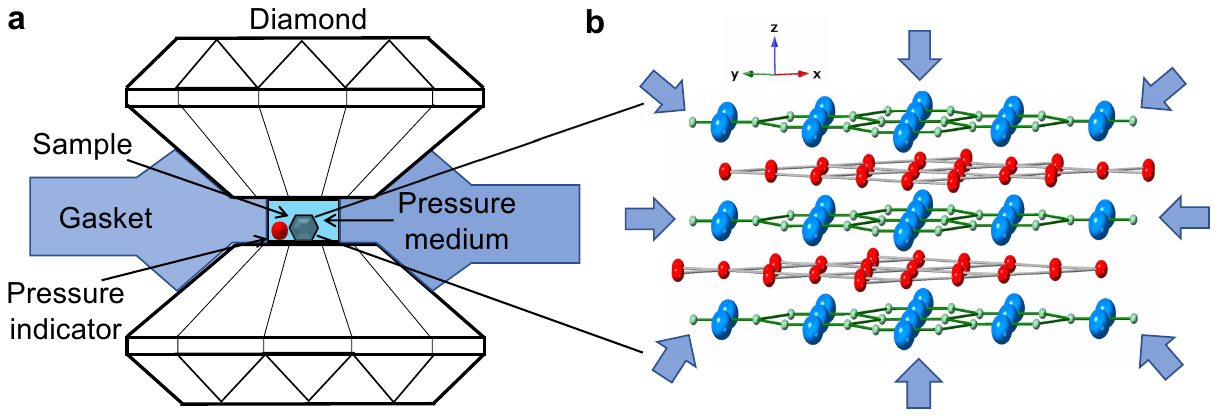}

    \vspace{-10.3cm}
    \includegraphics[width=1.33\textwidth]{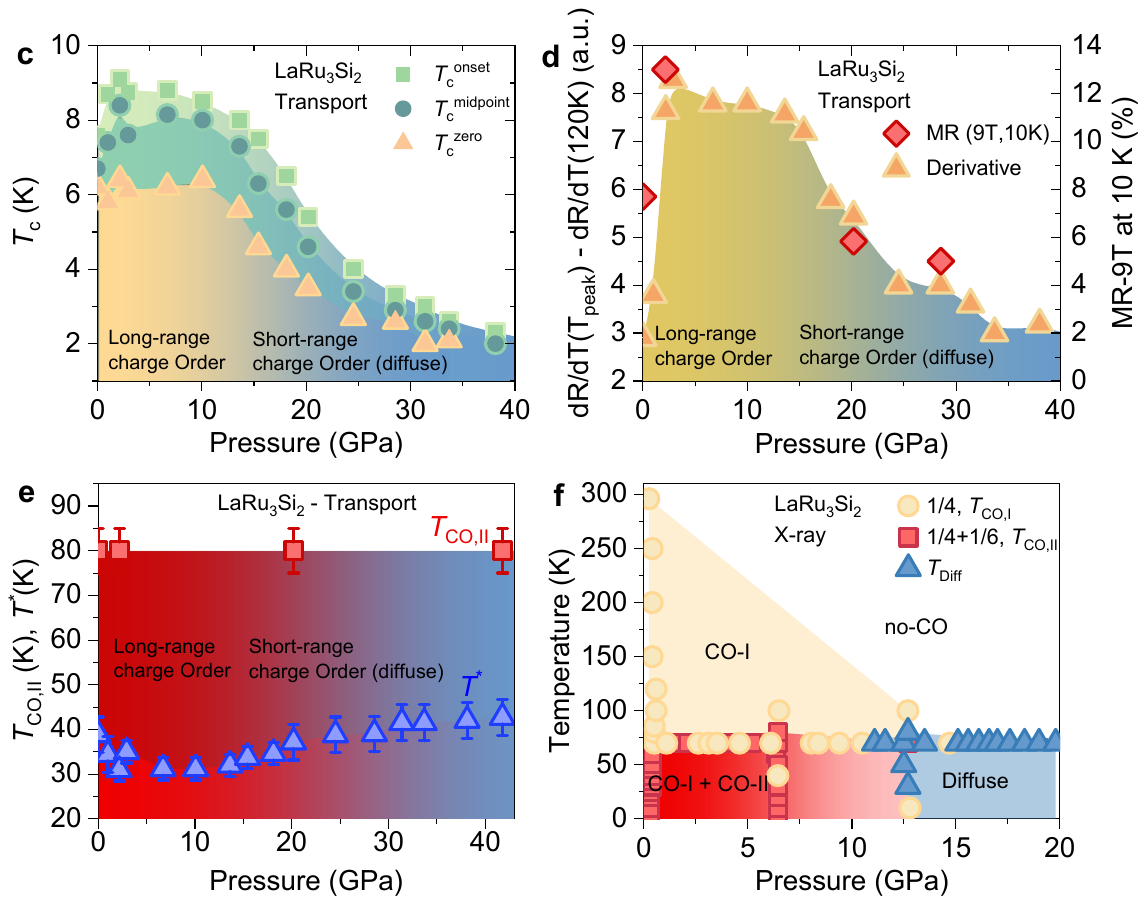} 
    \vspace{-2.5cm}
    \caption{\textbf{Temperature-Pressure Phase Diagrams.} 
    \textbf{(a)} Schematic illustration of the diamond anvil cell setup, showing the anvils, gasket, sample, pressure indicator, and pressure-transmitting medium.  \textbf{(b)} Side view of the atomic structure of LaRu$_{3}$Si$_{2}$. The Ru atoms construct a kagome lattice (red middle-size circles), while the Si (green small-size circles) and La atoms (blue large-size circles) form a honeycomb and triangular structure, respectively. \textbf{(c)} Pressure dependence of the superconducting critical temperature, determined from the temperature dependence of resistivity using three criteria: onset, midpoint, and zero-resistance values. \textbf{(d)} The strength of the anomaly in resistivity near $T^{*}$ $\simeq$ 30-40 K. The magnetoresistance at 9 T is also shown as a function of pressure on the right axis. \textbf{(e)} Pressure dependence of $T^{*}$ and $T_{\rm CO,II}$, determined from the resistivity measurements. Long-range and short-range charge-ordered regions of the phase diagram is noted based on X-ray diffraction experiments. \textbf{(f)} Pressure evolution of the charge orders ($\frac{1}{4}$,~0,~0) and ($\frac{1}{6} $,~0,~0), as determined from X-ray diffraction experiments.}
    \label{fig1}
\end{figure*}

Within this landscape, LaRu$_{3}$Si$_{2}$ \cite{Vandenberg,GuguchiaPRM,GuguchiaPlokhikh} stands out as a kagome superconductor with a relatively high superconducting critical temperature ($T_{\rm c}$ = 6.5 K) and a charge order (propagation vector of ($\frac{1}{4}$,~0,~0)) with an exceptionally high onset temperature, $T_{\rm co,I}$ ${\simeq}$ 400 K-the highest recorded for any kagome superconductor \cite{GuguchiaPlokhikh}. This high-temperature charge order distinguishes LaRu$_{3}$Si$_{2}$ from its counterparts, suggesting an unusual stability of ordered electronic phases even at elevated temperatures. In addition to this high-temperature charge order, we identified a second charge ordering transition with a propagation vector of ($\frac{1}{6}$,~0,~0) below $T_{\rm co,II}$ ${\simeq}$ 80 K \cite{GuguchiaarXiv}, coexisting with ($\frac{1}{4}$,~0,~0)) charge order. The secondary charge order aligns with a significant change in the electronic structure, as indicated by the emergence of magnetoresistance. The magnetoresistance begins to increase below $T_{\rm co,II}$ ${\simeq}$ 80 K, with a sharper rise observed below $T^{*}$ $\simeq$ 35 K \cite{GuguchiaarXiv}. Additionally, a sign reversal in the Hall effect occurs across $T^{*}$ \cite{GuguchiaarXiv}. These transitions at 80 K and 35 K are not purely electronic but are accompanied by the onset of magnetism—at 35 K in zero field and at 80 K under high magnetic field as revealed by muon-spin rotation ($\mu$SR) measurements. Consequently, three distinct temperature scales have been identified: $T_{\rm co,I}$ ${\simeq}$ 400 K, $T_{\rm co,II}$ ${\simeq}$ 80 K, and $T^{*}$ $\simeq$ 35 K. This sequence of charge ordering and magnetic transitions below 80 K suggests a coupling between the electronic and magnetic degrees of freedom, raising intriguing questions about the role of magnetism in the ground state of LaRu$_{3}$Si$_{2}$. Despite these fascinating properties, the precise nature of the interplay between superconductivity and the dual charge orders, as well as the magnetic phase, remains unclear, especially given the complex phase landscape introduced by the kagome lattice.

\begin{figure*}
    \centering
    \includegraphics[width=\textwidth]{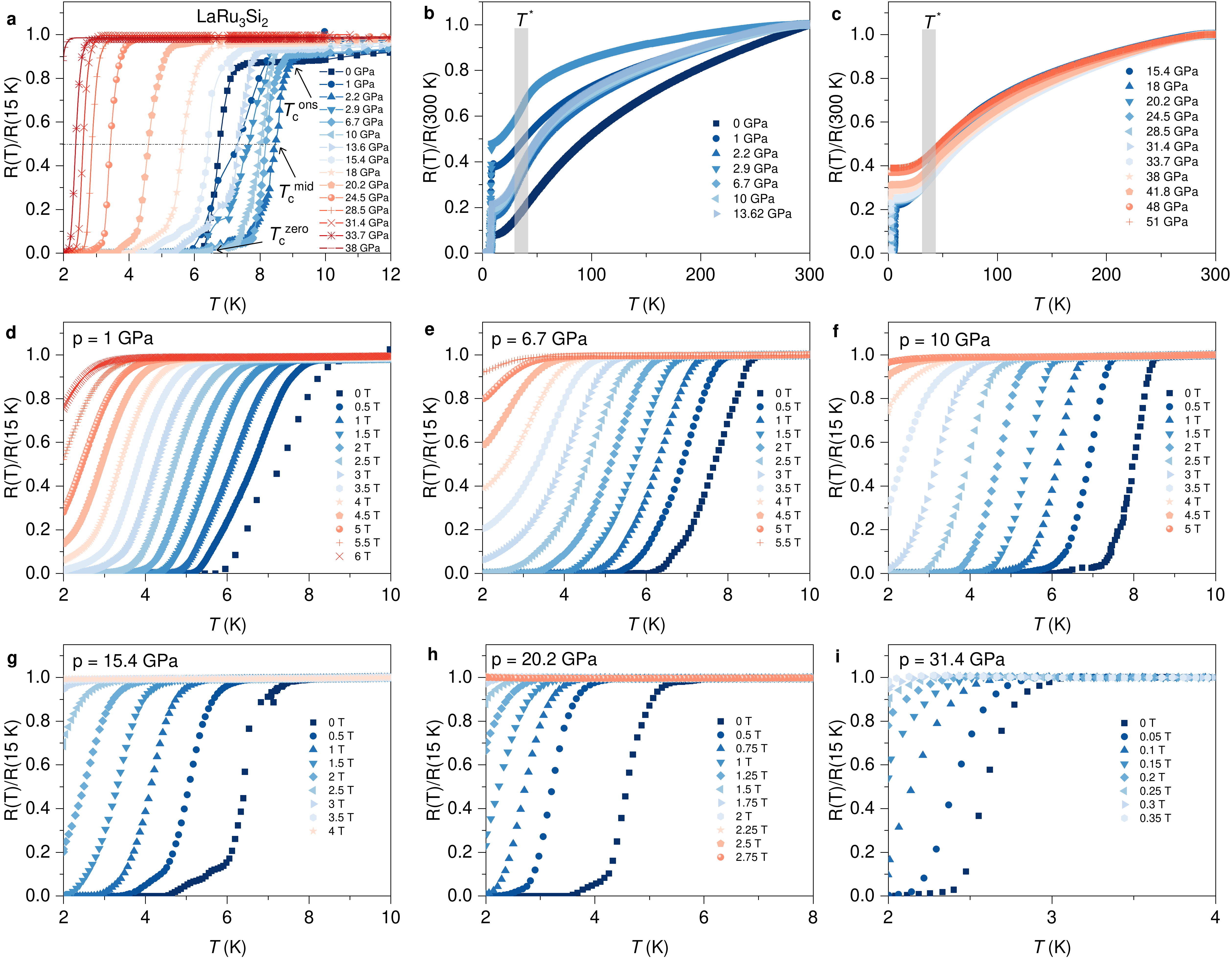}
    \vspace{0cm}
    \caption{\textbf{Pressure tuning of superconductivity.} \textbf{(a)} Temperature dependence of resistivity, normalized to the value at 15 K, focusing on the low-temperature region to highlight the superconducting transitions recorded at ambient pressure and under hydrostatic pressures up to 38 GPa. \textbf{(b-c)} Temperature dependence of resistivity, normalized to the value at 300 K, measured over a wide temperature range at various pressures within 0-10 GPa and 13.6-51 GPa. \textbf{(d-i)} Temperature dependence of resistivity below 10 K, measured under different applied magnetic fields at selected pressures: \textbf{(d)} 1 GPa, \textbf{(e)} 6.7 GPa, \textbf{(f)} 10 GPa, \textbf{(g)} 15.4 GPa, \textbf{(h)} 20.2 GPa, and \textbf{(i)} 31.4 GPa.}
    \label{fig1}
\end{figure*}

To investigate this interplay, we conducted a comprehensive study using a combination of resistivity, magnetoresistance, and single-crystal X-ray diffraction measurements under extreme hydrostatic pressures, reaching up to 40 GPa using a diamond-anvil cell (Figure 1a and b). High-pressure techniques offer a powerful tool to tune electronic and structural properties, allowing us to systematically map how superconductivity and charge orders respond to external perturbations in this kagome superconductor. By applying such pressures, we observe an evolution in $T_{\rm c}$, which initially increases from 6.5 K to 9 K at 2 GPa and remains steady up to 12 GPa, after which it gradually decreases, reaching 2 K at 40 GPa (Fig. 1c). This dome-shaped superconducting phase diagram is a hallmark of potential unconventional superconductivity. Notably, the strength of the resistivity anomaly across $T^{*}$ as well as the base-$T$ value of magnetoresistance exhibits a similar dome-shaped pressure dependence (Fig. 1d). Furthermore, the suppression of $T_{\rm c}$ beyond the superconducting dome correlates with a decrease in the onset temperature of the primary charge order under pressure. Long-range order persists up to 12 GPa, where the onset temperatures of the two charge orders converge $T_{\rm co,I}$ ${\simeq}$ $T_{\rm co,II}$ and transition into short-range order. These findings underscore two key points: first, distortions associated with charge orders play a critical role in enhancing the superconducting critical temperature in LaRu$_{3}$Si$_{2}$ and second, the superconducting $T_{\rm c}$ and $T^{*}$, along with their electronic and magnetic responses, are intimately connected.

\begin{figure}
    \centering
    \includegraphics[width=0.45\textwidth]{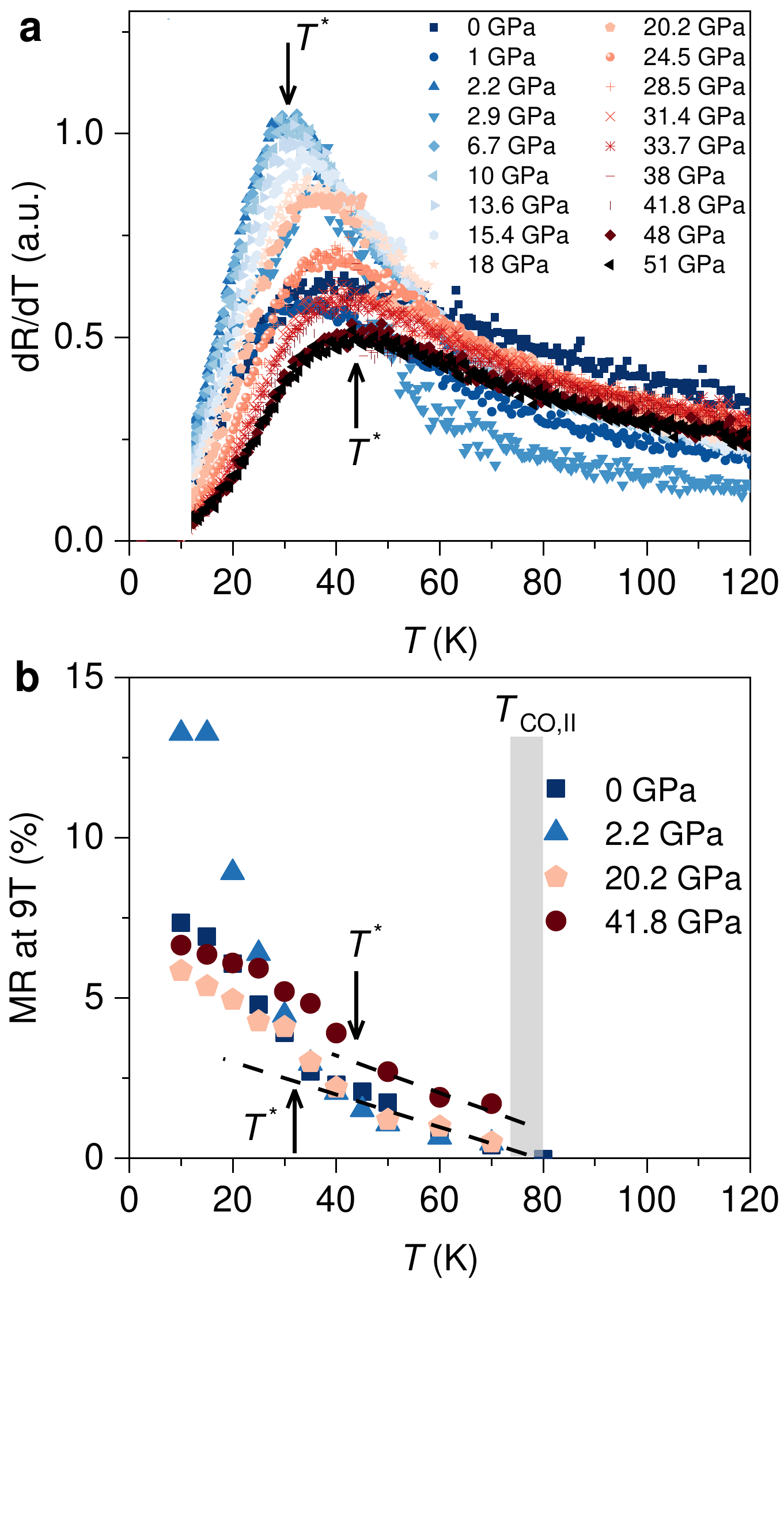}
    \vspace{-3cm}
    \caption{\textbf{Pressure tuning of normal state transport in LaRu$_{3}$Si$_{2}$.} \textbf{(a)} The temperature dependence of the derivative of resistivity dR/dT, measured under various hydrostatic pressures up to 51 GPa. \textbf{(b)} The temperature dependence of magnetoresistance, measured at selected pressures of $p$=0 GPa, 2.2 GPa, 20.2 GPa and 41.8 GPa.}
    \label{fig1}
\end{figure}

\section{Results}

Figure 2a presents the temperature dependence of resistivity for LaRu$_{3}$Si$_{2}$, focusing on the low-temperature range, measured at ambient pressure and under applied pressures up to 38 GPa. At ambient pressure, the system displays a well-defined superconducting transition, with an onset temperature $T_c^{\rm onset}$  of 6.5 K, and reaching zero resistance below 6 K, indicating a robust superconducting phase. With an applied pressure of 1 GPa, $T_{\rm c,onset}$ increases noticeably to 9 K, although the transition becomes relatively broader, suggesting pressure-induced modifications in the superconducting properties. As the pressure increases further, up to 10 GPa, $T_c^{\rm onset}$ remains at 9 K; however, the superconducting transition sharpens, indicating an enhancement in phase coherence or homogeneity in the superconducting state under moderate pressures. This stability in the superconducting onset up to 10 GPa reflects the systems resilience to pressure within this range. Beyond 10 GPa, the onset of superconductivity begins to decrease gradually, signifying a pressure-induced suppression of superconductivity. By 38 GPa, the onset temperature has decreased significantly, reaching 2 K, pointing to a diminishing superconducting state at high pressures. 
Figures 2d–i illustrate the suppression of the transitions under magnetic fields at selected hydrostatic pressures, confirming the superconducting nature of the transitions. Additionally, they demonstrate a decrease in the critical field as pressure is reduced.

For clarity in interpreting these observations, we define three characteristic temperatures: the onset temperature $T_c^{\rm onset}$, the midpoint temperature $T_c^{\rm midpoint}$ (where resistivity decreases by 50${\%}$), and the temperature $T_c^{\rm zero}$ below which resistivity reaches zero. These defined temperatures are illustrated in the phase diagram shown in Figure 1c, which captures the overall pressure-dependent superconducting behavior. The phase diagram exhibits a dome-shaped superconducting region, indicating an optimal pressure range where superconductivity is most robust, with suppression occurring beyond this optimal range. This dome-shaped behavior highlights the intricate interplay between pressure and superconductivity in LaRu$_{3}$Si$_{2}$, potentially revealing insights into the underlying mechanisms of superconductivity in this system.\\

\begin{figure*}
    \centering
    \includegraphics[width=\textwidth]{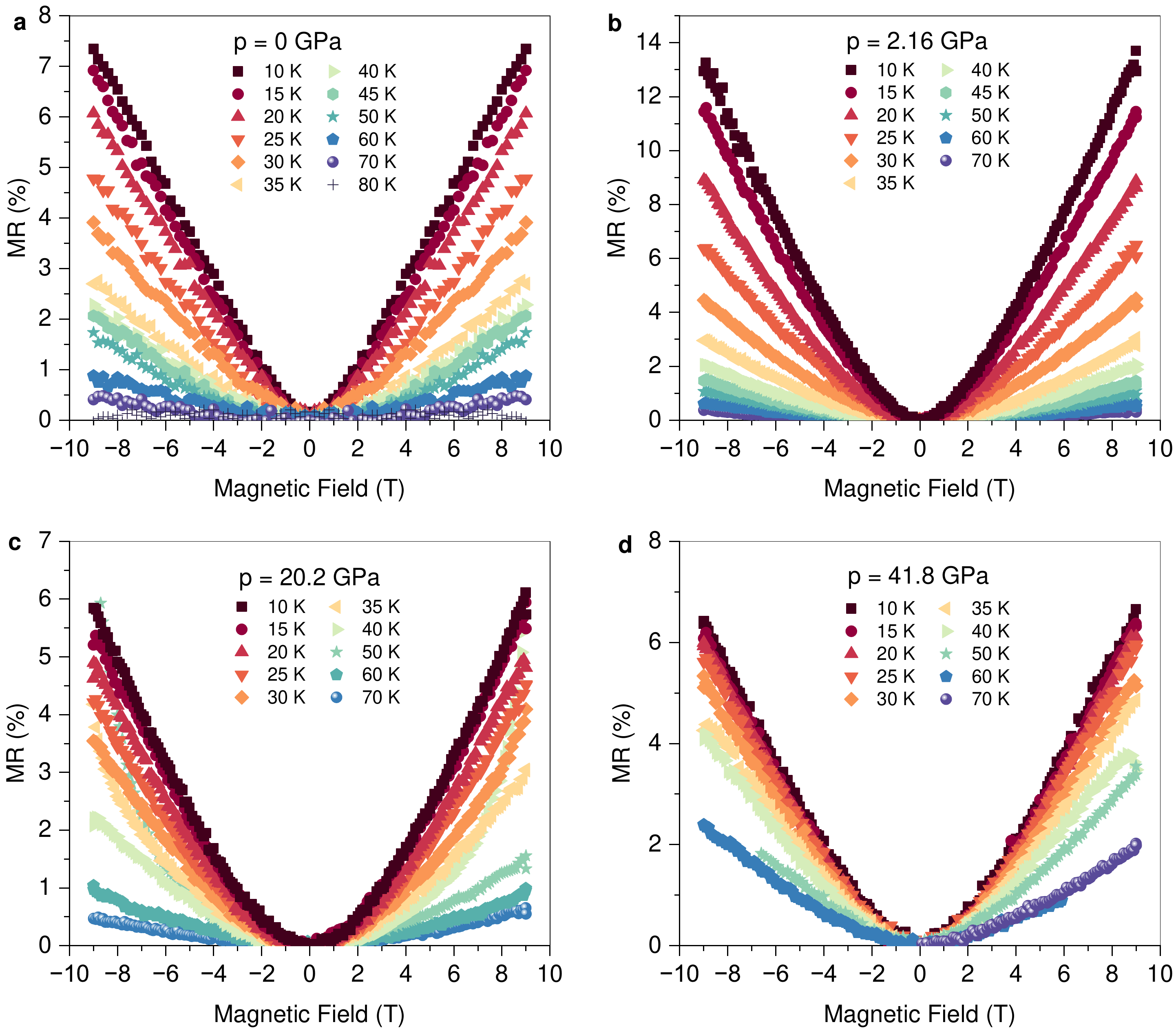}
    \vspace{0cm}
    \caption{\textbf{Pressure tuning of magnetotransport characteristics in LaRu$_{3}$Si$_{2}$.} The magnetoresistance measured at various temperatures across the charge ordering temperature $T_{\rm CO,II}$ ${\simeq}$ 80 K at ambient pressure \textbf{(a)} and under various hydrostatic pressures of  \textbf{(b)} $p$=2.2 GPa, \textbf{(c)} $p$ = 20.2 GPa and  \textbf{(d)} $p$ = 41.8 GPa.}
    \label{fig1}
\end{figure*}

\begin{figure*}
    \centering
    \includegraphics[width=\textwidth]{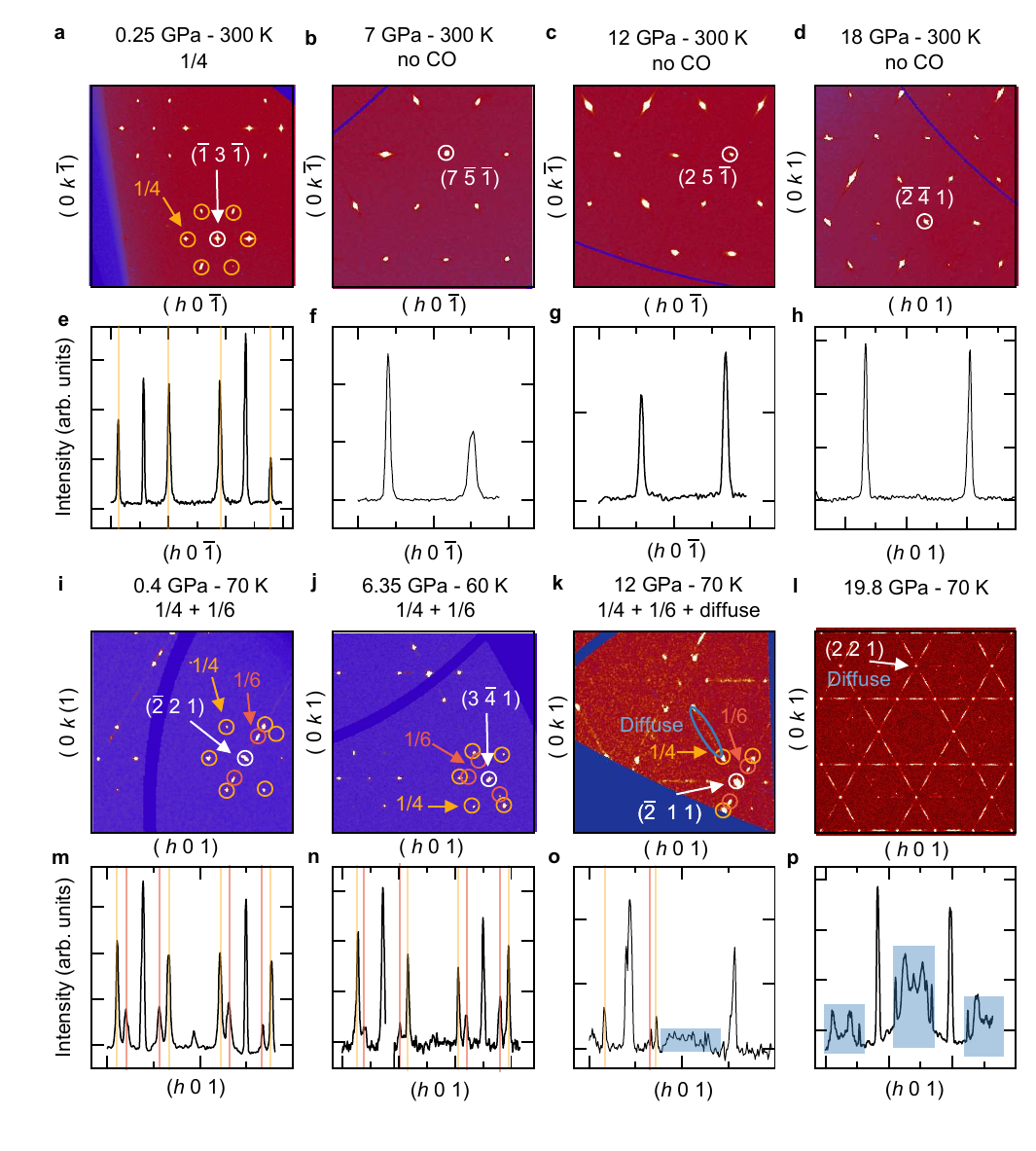}
    \vspace{-1.25cm}
    \caption{\textbf{Pressure tuning of charge orders in LaRu$_{3}$Si$_{2}$.} \textbf{(a-d)} Reconstructed reciprocal space along the (h k 1) direction, measured at $T$=300 K for various hydrostatic pressures  (a) $p$=0.15 GPa, (b) 7 GPa, (c) 12 GPa, and (d) 18 GPa. \textbf{(e-h)} Corresponding 2D cuts.
    Vertical lines in panel (e) indicate the charge order peaks with a propagation vector of ($\frac{1}{4}$,~0,~0). \textbf{(i-l)} Reconstructed reciprocal space along the (h k 1) direction, measured at $T$ ${\simeq}$ 70 K for various hydrostatic pressures $p$=0.4 GPa (i), 6.35 GPa (j), 12 GPa (k), and 19.8 GPa (l). \textbf{(m-p)} Corresponding 2D cuts, shown on a logarithmic scale. Circles indicate the charge order peaks with a propagation vectors of ($\frac{1}{4}$,~0,~0) and ($\frac{1}{6}$,~0,~0). Vertical orange and red lines indicate the charge order peaks with a propagation vectors of ($\frac{1}{4}$,~0,~0) and ($\frac{1}{6}$,~0,~0), respectively.}
    \label{fig1}
\end{figure*}

Next, we discuss the normal-state response observed in the resistivity data. Notably, no anomaly is detected across $T_{\rm co,II}$ either at ambient pressure or under applied pressure. However, an anomaly is evident across $T^{*}$, which is weak at ambient pressure but becomes clear in the first derivative of the resistivity, as shown in Figure 3a. Interestingly, applying pressure enhances the prominence of this anomaly, particularly in the pressure range between 2.2 GPa and 15.4 GPa. In this range, the anomaly is clearly visible in the resistivity data and is further accentuated in the first derivative, where sharper and more intense peaks are observed. Above 15.4 GPa, the anomaly progressively weakens, and the peaks in the first derivative become less intense and broader. To characterize the pressure evolution of this anomaly, we calculated the difference in the first derivative d$R$/d$T$ between its maximum value at $T^{*}$ and its value at 120 K. This difference serves as a measure of the anomaly's strength and is plotted as a function of pressure in Fig. 1d. Remarkably, the anomaly exhibits a dome-shaped pressure dependence, with its maximum occurring in the pressure range of 1–15.4 GPa, coinciding with the pressure range where the superconducting transition temperature $T_{\rm c}$ reaches its peak. These results suggest that the anomaly across $T^{*}$ becomes most pronounced when superconductivity is optimal. In Fig. 1e, the pressure dependence of $T^{*}$, determined as the peak position in the first derivative, is presented. Interestingly, $T^{*}$ exhibits a slight dip around 10 GPa, coinciding with the region of maximum $T_{\rm c}$ in the phase diagram.

Both temperature scales, $T^{*}$ and $T_{\rm co,II}$, in LaRu$_{3}$Si$_{2}$, are also reflected in magnetoresistance (MR) measurements. Magnetotransport techniques \cite{Giraldo,Novak,XWei,LDas,Wang2005,TailleferNature,TailleferPRB}, known for their sensitivity to charge-order transitions, utilize MR as an indicator of the mean free path integrated over the Fermi surface \cite{LDas}. This approach is particularly effective in detecting changes in scattering anisotropy and Fermi surface reconstructions. Figures 4a–d present the MR in LaRu$_{3}$Si$_{2}$ under a perpendicular magnetic field across the temperature range of 10 K to 80 K, measured at $p$ = 0 GPa and at selected pressures of $p$ = 2.2 GPa, 20.2 GPa, and 41.8 GPa. Additionally, Figure 3b shows the temperature dependence of MR at 9 T, recorded under various pressures. Within the primary 1/4 charge-ordered state, the MR remains negligible and only begins to appear below the 1/6 charge ordering temperature $T_{\rm co,II}$. At $p$ = 0 GPa, the MR starts to increase at $T_{\rm co,II}$, with a steeper rise occurring below $T^{*}$ (see Figs. 3b and 4a). Under a pressure of 2.2 GPa, both the onset of MR and its base-temperature value at 9 T remain unchanged down to $T^{*}$, below which the MR increases significantly (see Figs. 3b and 4b). Notably, the MR(9T) value at 10 K reaches 14 $\%$, nearly twice the value observed at ambient pressure. However, when pressure exceeds 15 GPa, the MR decreases and eventually becomes smaller than the value recorded at ambient pressure (see Figs. 4c and 4d). Despite these changes, the onset of charge order at $T_{\rm co,II}$ remains nearly unaffected. These MR experiments reveal that the absolute value of MR is maximized within the pressure range where superconductivity is optimal, highlighting a strong connection between charge order, magnetotransport properties, and superconductivity in LaRu$_{3}$Si$_{2}$. 

To complement the resistivity experiments, we conducted X-ray diffraction measurements up to 20 GPa, providing direct insight into both the 1/4 and 1/6 charge orders. In Figures 5$\bf{a-d}$, we present reconstructed reciprocal-space patterns along the (hk1) direction for 300 K at selected pressures for a single crystal of LaRu$_{3}$Si$_{2}$. Figures 5e-h shows the corresponding 2D cuts. At higher temperatures $T$ = 300 K (Fig. 5a and e), the diffraction pattern reveals fundamental Bragg peaks ${\tau}$ and superlattice peaks at $Q$ = ${\tau}$ + $q_{i}$ with $q_{1}$ = ($\frac{1}{4}$,0,0) and $q_{2}$ = (0,$\frac{1}{4}$,0). The application of pressure suppresses the onset temperature of the primary charge order ($T_{\rm co,I}$) such that already at 7 GPa charge order peaks are not visible (Fig. 5b-d, 5f-h).
As the temperature decreases below $T_{\rm co,II}$, approximately 80 K (see Fig. 5i-l), an additional set of reflections emerges at positions corresponding to $q_{1}'$=($\frac{1}{6},0,0$), $q_{2}'$=(0,~$\frac{1}{6},0$), and $q_{3}'$=($\frac{1}{6},\frac{-1}{6},~0$). Significantly, both $\frac{1}{4}$ and $\frac{1}{6}$ charge orders coexist below 80 K, persisting into the superconducting state. While the application of pressure suppresses the onset temperature of the primary charge order ($T_{\rm co,I}$), the onset temperature ($T_{\rm co,II}$) below which 1/4 and 1/6 charge orders coexist remains nearly unaffected (see Figs. 5i-k, 5m-o). At approximately 12.5 GPa, $T_{\rm co,I}$ and $T_{\rm co,II}$ converge, beyond which no well-defined superlattice peaks are observed. Instead, broad, diffuse scattering intensity emerges at the same onset temperature ($T_{\rm co,II}$) (see Figs. 5l-p), but it does not coalesce into sharp Bragg diffraction peaks down to base temperature. Notably, the transition from long-range charge order to a short-range state at 12.5 GPa coincides with the pressure at which the superconducting transition temperature begins to decrease. This correlation suggests a positive relationship between charge order and superconductivity in LaRu$_{3}$Si$_{2}$.

\section{Discussion}

Charge order is a common feature in materials with strong electronic correlations and significant electron–phonon interactions. It has been observed in various systems, including cuprate high-temperature superconductors, colossal magnetoresistive manganites, transition-metal dichalcogenides (TMDs), and more recently, kagome lattice metals. Charge order is often linked to unconventional quantum phases, underscoring its critical role in understanding and manipulating novel states of matter. One of the central questions in condensed matter physics is the interplay between charge order, normal-state electronic and magnetic responses, and superconductivity.
In many unconventional superconductors, including the recently discovered $A$V$_{3}$Sb$_{5}$ kagome systems, superconductivity and charge order are found to compete. However, in this work, we demonstrate a different behavior in LaRu$_{3}$Si$_{2}$: a dome-shaped superconducting phase diagram is observed within a broad pressure range of 2–12.5 GPa, with a maximum superconducting transition temperature ($T_{\rm c}$) as high as 9 K, marking a record $T_{\rm c}$ for this system. Notably, our findings reveal not a competition but rather a coexistence between superconductivity and charge order. The superconducting critical temperature is optimized when the normal-state electronic response, magnetoresistance, and charge order are also optimized.

Previously, we showed that in the charge-ordered phase, there is a significant disproportionation of Ru–Ru bonds, resulting in the formation of distinct short and long bonds, indicative of bond or charge ordering. This ordering arises from distortions in Ru–Ru bonds, which are associated with the dz$^{2}$ orbitals of Ru atoms, forming a unique kagome band structure. This connection suggests a potential link between charge ordering and the underlying physics of the kagome lattice. In this work, we demonstrate that the high superconducting transition temperature occurs when the charge order is optimal. Specifically, the out-of-plane Ru–Ru distortions and the resulting changes in electronic responses favor a higher $T_{\rm c}$. This insight opens exciting opportunities to further enhance $T_{\rm c}$ by applying in-plane uniaxial strain to distort the Ru lattice, providing a promising avenue for future research.

\section{Conclusion}

In this work, we employ a combination of resistivity, magnetoresistance, and single-crystal X-ray diffraction techniques to probe the superconductivity and charge orders in LaRu$_{3}$Si$_{2}$ under hydrostatic pressures up to 40 GPa. This high-pressure approach enables us to map out the response of superconductivity, normal state transport and charge orders to extreme conditions, offering new insights into their interplay and the broader physics of kagome superconductors.

Namely, our experiments reveal LaRu$_{3}$Si$_{2}$ as a unique member of the kagome superconductor family, where superconductivity, normal state and charge order not only coexist but are strongly connected. In LaRu$_{3}$Si$_{2}$, the relationship is notably synergistic: lattice distortions from charge order may enhance superconductivity, with $T_{\rm c}$ reaching its peak when the normal state transport response and the charge order is optimal. This work establishes the system LaRu$_{3}$Si$_{2}$ as a valuable model for theoretical exploration into the mechanisms behind superconductivity in kagome materials. By uncovering an intimate connection of kagome lattice superconductivity with both charge order and related electronic responses, this study deepens our understanding of kagome lattices as platforms for novel quantum phenomena and aids in the search for new superconductors with intertwined orders and tunable ground states.

\section{Methods}
\subsection{High-pressure resistivity experiments}

For high-pressure experiments, a screw-clamped diamond-anvil cell (DAC) manufactured from the nonmagnetic alloy MP35N and equipped with Boehler-Almax design diamond anvils with 500-${\mu}$m culets was used. The tungsten gasket was insulated with a cubic BN-epoxy mixture. A single crystal sample of suitable size (120 ${\mu}$m ${\times}$ 120 ${\mu}$m ${\times}$ 10 ${\mu}$m) was cut and placed into the central hole of the gasket filled with NaCl as a pressure-transmitting medium. The electrical leads were fabricated from 5-${\mu}$m-thick Pt foil and attached to the sample in a van der Pauw conﬁguration. Electrical resistivity was measured at different pressures in the temperature range 1.8–300 K in a magnetic field up to 9 T using the electrical transport option of a Physical Property Measurement System (PPMS-9, Quantum Design). The pressure was determined using the ruby scale by measuring the luminescence from the small chips of ruby placed in contact with the sample.\\

\subsection{High-pressure X-ray Diffraction experiments}

X-ray diffraction was performed at ID27 \cite{Mezouar},  ESRF using monochromatic X-rays with a wavelength of 0.3738 ${\AA}$ and a spotsize of 0.6 ${\times}$ 0.6 ${\mu}$m$^{2}$. The sample was mounted on a membrane driven diamond anvil cell with 70 degrees angular aperture designed at ESRF to reach hydrostatic pressures up to 20 GPa. The pressure cell is made of CuNi$_{2}$Be. The diamonds were single crystals with Boehler-Almax design and culet sizes of 600 $\mu$m. Stainless steel was used as gasket material. Helium was used as the pressure transmitting medium and the applied pressure was measured by the fluorescence of Ruby. The diamond anvil cell was mounted into a Helium flow cryostat (ESRF). The applied pressures were kept constant during cooling by adapting the membrane pressure. X-ray diffraction was collected with a Eiger2 CdTe detector (DECTRIS AG, Baden-Daettwil, Switzerland) in shutterless mode and continuous rotation over 64 degrees with readout every 0.1 degree. CrysAlisPro (Rigaku) was used for data reduction and reciprocal space reconstructions.\\

\section{Acknowledgments}~
Z.G. acknowledges support from the Swiss National Science Foundation (SNSF) through SNSF Starting Grant (No. TMSGI2${\_}$211750). Z.G. acknowledges usful discussions with Prof. T. Neupert and Dr. M.H.~Fischer. I.P. acknowledges support from Paul Scherrer Institute research grant No. 2021${\_}$0134. We acknowledge the ESRF for provision of beamtime within the in-house proposal IH-HC-3952, Jeroen Jacobs for preparation and gas loading of the high-pressure cells, Gaston Garbarino for help with the high-pressure cryostat and Mohamed Mezouar with help aligning the beamline.\\

\section{Author contributions}~
Z.G. conceived and designed the project. Z.G., S.M. and D.J.G. supervised the project. Crystal growth: H.N. and S.N.. High pressure transport experiments, analysis and corresponding discussions: K.M., S.M., and Z.G.. High pressure X-ray diffraction experiments at ESRF, analysis and corresponding discussions: I.P., C.M.III, J.N.G., V.S., D.J.G., B.W., and Z.G. Figure development: K.M., Z.G., S.M., and J.N.G.. Writing of the paper: Z.G. with contributions from all authors. All authors discussed the results, interpretation, and conclusion.  

\section{Data availability}~
The high-pressure X-ray diffraction data is available at https://doi.esrf.fr/10.15151/ESRF-ES-1402335301. High pressure transport data are available from authors upon request.\\

\end{document}